\documentclass{appolb}
\usepackage{epsfig}
\begin{document}
%\date{\today}
%\pagestyle{plain}
%% uncomment the following line to get equations numbered by (sec.num)
%\eqsec
%\newcount\eLiNe\eLiNe=\inputlineno\advance\eLiNe by -1
\title{Study of hyperon-hyperon correlations and search for $H$-dibaryon using STAR detector at RHIC%
%\thanks{Send any remarks to {\tt neha@physics.ucla.edu}}%
\thanks{Presented at the conference Strangeness in Quark Matter 2011, Cracow, Poland, September 18-24, 2011. }%
}
\author{Neha Shah for the STAR collaboration
\address{Department of Physics and Astronomy, University of California, Los Angeles, California - 90045, U.S.A.}}
\maketitle

\begin{abstract}

High yield of strange particle production per central collision in nucleus-nucleus collisions at RHIC and high statistics data set from STAR experiment allow us to measure $\Lambda\Lambda$ correlations. The $\Lambda\Lambda$ correlation measurement is also closely related to a possible six-quark state $H$-dibaryon, which is yet to be observed experimentally. The first measurement of $\Lambda\Lambda$ correlations in Au+Au collisions for $\sqrt{s_{NN}}= 39$, 62.4 and 200 GeV  and search for $H$-dibaryon with its weak decay $\Lambda p\pi$  in Au+Au collisions for $\sqrt{s_{NN}}= 200$ GeV using the STAR experiment at RHIC are presented here. 

\end{abstract}


\section{Introduction}

The production of large number of multi-strange hyperons per central nucleus-nucleus collision at RHIC~\cite{strangebaryon} allows us to study hyperon-hyperon interactions through measurement of particle correlations and search for exotic particles like dibaryons. The knowledge of hyperon-hyperon interaction is important for understanding the baryon-baryon interactions in a unified way. Further the study of hyperon-hyperon interactions is also associated with a possible six quark state $H$-dibaryon.

In 1977 Jaffe~\cite{Jaffe} predicted a six quark state, $H$-dibaryon, with hypercharge ($Y$) = 0 and strangeness ($S$) = 2 to be stable against strong decay, but not to weak decay. Recent lattice QCD calculations from HAL~\cite{HALQCD} and NPLQCD~\cite{NPLQCD} collaborations indicate existence of a bound $H$ for pion mass above the physical mass. The coalescence model predicted $H$-dibaryon production rate ranges from 10$^{-2}$ - 10$^{-4}$~\cite{dover,Lie} for Au+Au collisions at $\sqrt{s}=200$ GeV.  It has been proposed that the $H$ would appear as a bump in the $\Lambda\Lambda$ invariant mass spectra if $H$  is a resonance state, or it would lead to a depletion of the $\Lambda\Lambda$ correlation near the threshold if $H$ is weakly bound, which can be used to probe whether there is a stable $H$ or resonance. In addition to $H$-dibaryon there are other six quark states with higher strangeness content are predicted: $\Xi\Xi$ ($S$ = 4)~\cite{Miller}, $\Omega\Omega$ ($S$=6)~\cite{zhang}. Various experiments have explored the $\Lambda\Lambda$~\cite{KEK}, $\Sigma p$~\cite{E910} and $\Lambda p\pi$~\cite{E910,kTeV} invariant mass spectrum to look for $H$-dibaryon signal, but statistics were too low to arrive at any conclusion about its existence.

In heavy ion collisions hyperon-hyperon interactions can be studied by measurement of the two hyperon correlation functions. Two particle correlation functions provide a powerful tool to extract dimensions of the emitting source from which these particles are produced . If the hyperons are produced in early stage of fireball expansion, they will be localized near the center of the system and the measured dimension of the system would be 2-3 fm. However the measurement of hyperon-hyperon correlation function is challenging because of their low production cross-section. Recently STAR has collected unprecedented high statistics data for Au+Au collision at RHIC, which provides a unique opportunity to look for the exotic particles, like $H$-dibaryon and $\Lambda\Lambda$ correlations.

\section{Analysis details}

\subsection{$\Lambda$ identification}

\begin{figure}[hbt]
\begin{center}
\epsfxsize = 3.4in
\epsfysize = 2.9in
\epsffile{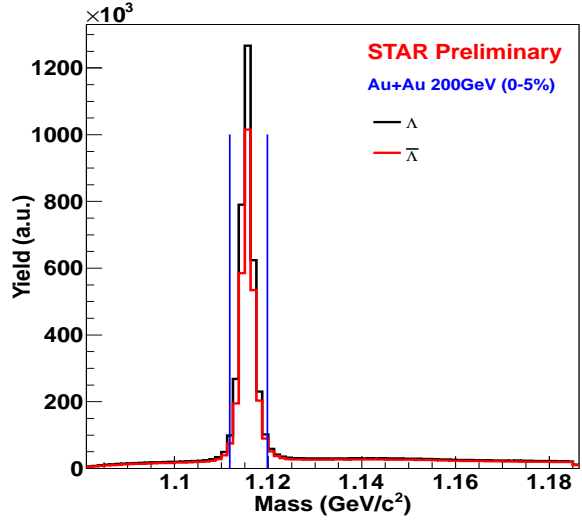} 
\end{center}
\caption{The invariant mass distribution for $\Lambda$($\bar{\Lambda}$) from 0-5$\%$ most central Au+Au collisions at $\sqrt{s_{NN}}=200$ GeV. The $\Lambda$($\bar{\Lambda}$) candidates falling in the mass range 1.112 to 1.12 GeV/$c^{2}$, shown by blue lines, were selected for the correlation measurement.}
\label{fig:IM}
\end{figure}

For this analysis, the data from the Au+Au collisions at $\sqrt{s_{NN}}=39$ and  62.4 GeV taken during the beam energy scan program, and data from the Au+Au collisions at $\sqrt{s_{NN}}=200$ GeV using STAR detector at RHIC were used. The total number of events analyzed were about 149.8 million for 39 GeV, 75.6 million for 62.4 GeV and 275.5 million for 200 GeV. $\Lambda$ identification and momentum determination were carried out with Time Projection Chamber (TPC). The decay channel $\Lambda \rightarrow p\pi$ with branching ratio 63$\%$ was used for particle identification. Protons and pions are identified by using specific ionization energy loss ($dE/dx$) in the TPC gas. The $\Lambda$ candidates were formed out of a pair of positive and negative tracks whose trajectories point to a common secondary decay vertex which was well separated from the primary vertex. All candidates having invariant mass from 1.112-1.12 GeV/$c^{2}$ have been considered. The decay length of $\Lambda$ candidate was asked to be more than 5 cm and the distance of closest approach from the primary vertex was taken to be less than 0.4 cm. Based on this conditions,  the invariant mass distribution of the $\Lambda$($\bar{\Lambda}$) in 0-5$\%$ centrality is shown in figure~\ref{fig:IM}, where black line(red line) is for $\Lambda$($\bar{\Lambda}$) candidates, with excellent signal to background ratio.

\subsection{Correlation Functions}

The two particle correlation function (CF) is defined as 

\label{myeq}
\begin{equation}
C(Q) = \frac{A(Q)}{B(Q)},
\end{equation}

\noindent
where A($Q$) is the distribution of the invariant relative momentum $Q=\sqrt{-q^{\mu}q_{\mu}}$, $q^{\mu}=p^{\mu}_{1}-p^{\mu}_{2}$, for a pair of $\Lambda(\bar{\Lambda})$ from the same event. B($Q$) is the reference distribution generated  by mixing particles from different events with approximately same vertex position along z-direction ($V_{z}$). The individual $\Lambda$ for a given mixed pair are required to pass the same single particle cuts applied to those that goes into the real pairs. 
The correlation between real $\Lambda$ and a fake $\Lambda$ candidate reconstructed from a pair that shares a daughter with the real $\Lambda$ candidate was rejected by removing pair of $\Lambda$ with same daughters. The effect from splitting of daughter tracks  is explored by looking at the angular correlation between the normal vectors to the decay planes of the $\Lambda$ in a given pair. No significant enhancement at very small angles was observed indicating no significant problem from track splitting. The mixed pairs are also required to satisfy the same pairwise cuts applied to the real pairs from one event. The efficiency and acceptance effects cancel out in the ratio $\frac{A(Q)}{B(Q)}$.

Corrections to the raw correlation functions are applied according to the expression

\label{myeq2}
\begin{equation}
C_{corrected}(Q) = \frac{C_{meassured}(Q)-1}{PairPurity(Q)}+1, 
\end{equation}
\noindent 
where the pair purity was calculated as the product of the signal ($s$) to signal plus background ($s+b$) ratios of the two $\Lambda$ of the pair $(i,j)$

\label{myeq3}
\begin{equation}
PairPurity(Q) = \frac{s}{s+b}(p_{ti})\times\frac{s}{s+b}(p_{tj}).
\end{equation}

The pair purity is constant over the range of invariant relative momentum ($Q$).

\begin{figure}[h]
\begin{center}
$\begin{array}{c@{\hspace{0.01in}}c}
\multicolumn{1}{l}{\mbox{}} &
\multicolumn{1}{l}{\mbox{}} \\ [-0.01cm]
\epsfxsize = 2.5in
\epsfysize = 2.8in
\epsffile{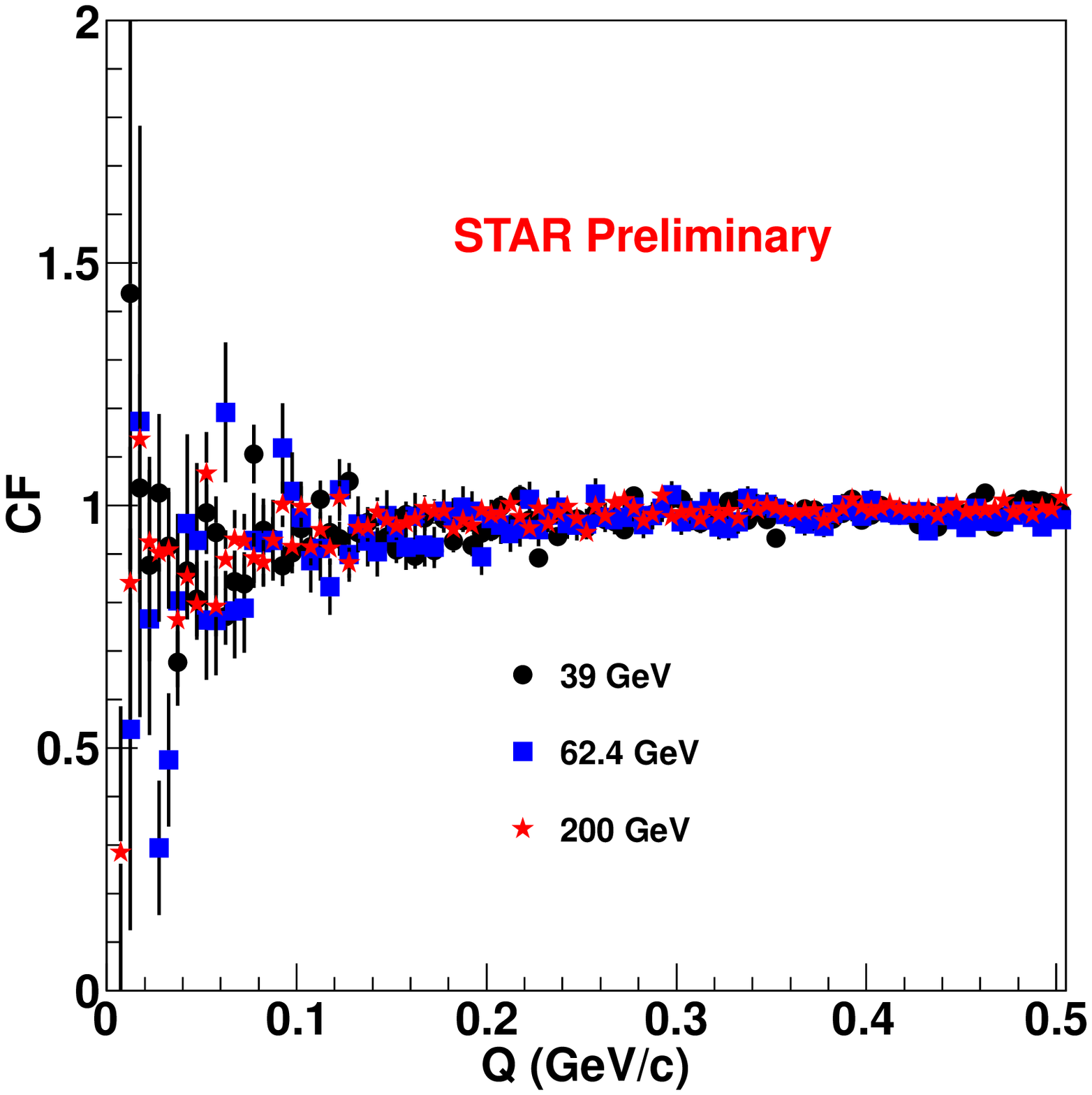} &
\epsfxsize=2.5in
\epsfysize = 2.8in
\epsffile{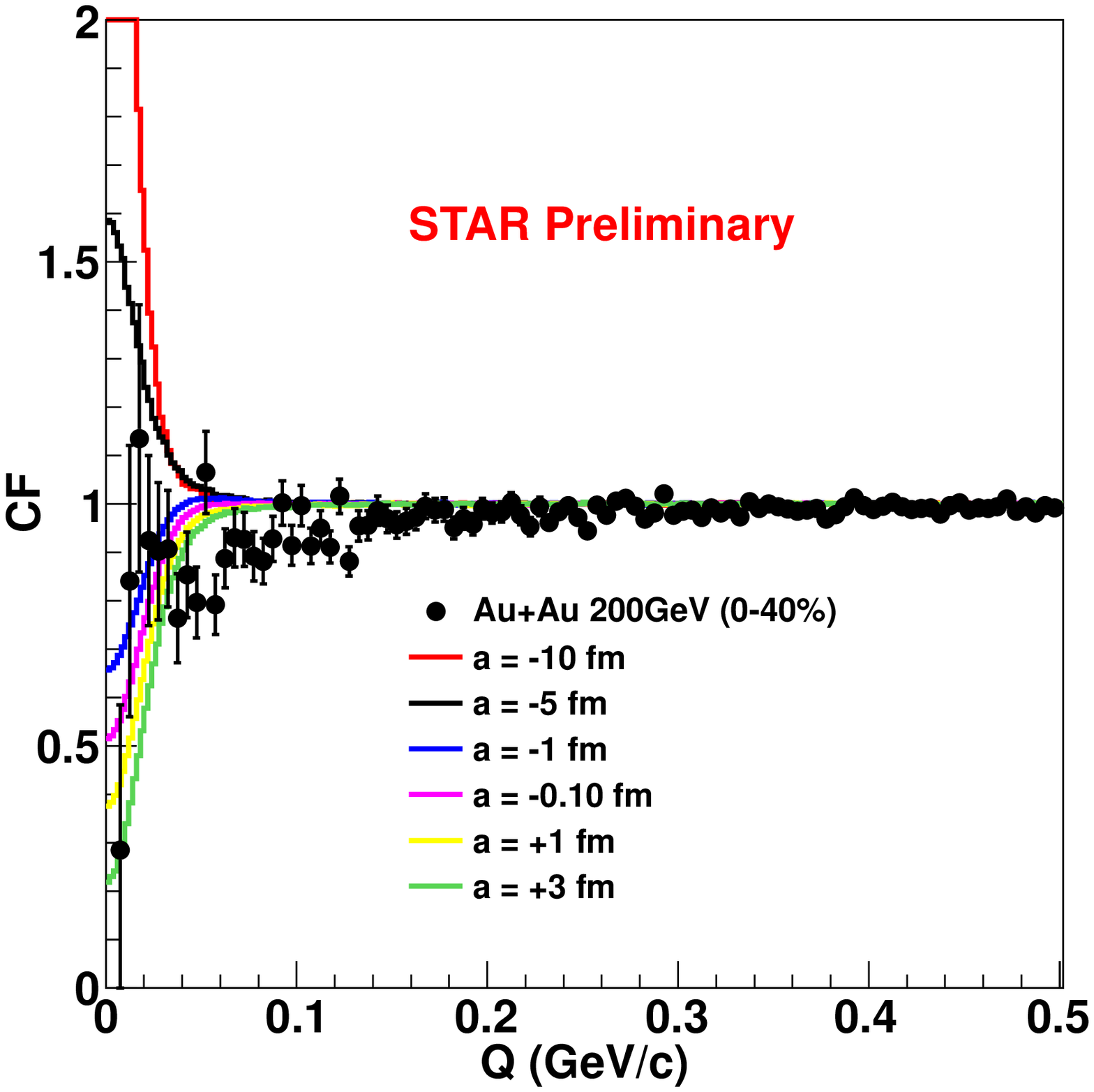} \\[0.2cm]
\mbox{\bf (A)} & \mbox{\bf (B)}
\end{array}$
\end{center}
\caption{(A) The purity corrected $\Lambda\Lambda$ correlation function from 0-40$\%$ most central Au+Au collisions at $\sqrt{s_{NN}}=39$ (circle), 62.4 (square) and 200 (star) GeV. (B) The $\Lambda\Lambda$ correlation function from 0-40$\%$ most central Au+Au collisions at $\sqrt{s_{NN}}=200$ GeV with theoretical correlation function for a Gaussian source of size $R_{x}=R_{y}=R_{z}=4$ fm for different scattering length (a).}
\label{fig:CF}
\end{figure}  

\subsection{$H$-dibaryon search}

To look for the $H$-dibaryon signal we have used one of the weak decay $H \rightarrow \Lambda p\pi$. The vertex of the $p\pi$ pair from sample closest to the target was taken to be the hypothetical $H$ vertex. The other $p\pi$ pair is required to be the hypothetical $\Lambda$, which decays at least 3.5 cm ($d_{\Lambda}$) away from the $H$ vertex and its mass is in the range 1.112-1.12 GeV/$c^{2}$. The other $p\pi$ pair mass was required to be less than 1.110 GeV/$c^{2}$. To connect  $\Lambda$ vertex to $H$ decay vertex, the angle between the line defined by $H$ and $\Lambda$ vertices and $\Lambda$ direction ($\theta_{\Lambda}$) is constrained to be smaller than 3 deg. Similar cut is applied to connect $H$ decay vertex with primary vertex: $\theta_{H} <$ 3 deg.

\section{Results}

Figure~\ref{fig:CF}(A) shows the experimental $\Lambda\Lambda$ correlation function for 0-40$\%$ most central Au+Au collisions at $\sqrt{s_{NN}}=39$, 62.4 and 200 GeV.  Figure~\ref{fig:CF}(B) shows theoretical correlation function for a Gaussian source of size $R_{x}=R_{y}=R_{z}=4$ fm for different scattering length(a)~\cite{Scott}. 
\begin{figure}[h]
\begin{center}
\epsfxsize = 3.4in
\epsfysize = 3.in
\epsffile{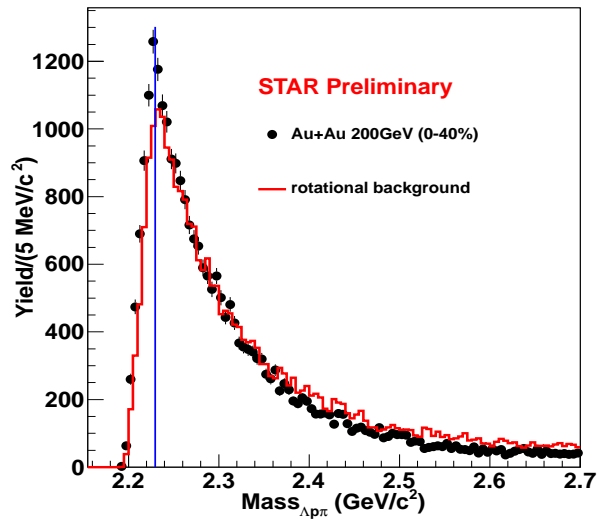} 
\end{center}
\caption{The invariant mass distribution of the $\Lambda p\pi$ from 0-40$\%$ most central Au+Au collisions at $\sqrt{s_{NN}}=200$ GeV obtained after applying cuts discussed for $H\rightarrow\Lambda p\pi$ analysis. Red line is the rotational background and blue line shows the $\Lambda\Lambda$ mass threshold.}
\label{fig:lmppi}
\end{figure}
For $\sqrt{s_{NN}}=39$ and 62.4 GeV the statistics are limited at low Q, hence correlation function for only 0-40$\%$ most central Au+Au collisions at $\sqrt{s_{NN}}= 200$ GeV is shown in figure~\ref{fig:CF}(B). However we require more statistics to extract scattering length and size of the emitting source from the data.

Figure~\ref{fig:lmppi} shows the $\Lambda p\pi$ invariant mass spectrum using the conditions mentioned above, where dots are the data points and the solid line is the background generated using the rotation of daughter tracks of the $\Lambda$ candidate. The data shown here are in very preliminary stage and more detailed analysis is required to arrive at any conclusion about the existence of $H$-dibaryon.

\section{Summary}

In summary, the first measurement of $\Lambda\Lambda$ correlation function in Au+Au collision at three different beam energies ($\sqrt{s_{NN}}=39$, 62.4 and 200 GeV) were presented. We also presented measurement of $\Lambda p\pi$ mass spectrum in Au+Au collisions at $\sqrt{s_{NN}}=200$ GeV to look for $H$-dibaryon signal.

\end{document}